\documentclass{PoS}

\title{Excited $D$ and $D_{s}$ meson spectroscopy from lattice QCD}

\ShortTitle{Excited $D$ and $D_{s}$ meson spectroscopy from lattice QCD}

\author{\speaker{Graham Moir}, Michael Peardon, Sin\'{e}ad M. Ryan, Christopher E. Thomas
        \\
        School of Mathematics, Trinity College, Dublin 2, Ireland\\
        E-mail: \email{moirg@tcd.ie}
\\
\\
(For the Hadron Spectrum Collaboration)
}

\abstract{We present highly excited spectra of charm-light and charm-strange mesons from 
dynamical lattice QCD. Our calculations are performed on anisotropic $N_{f} = 2 + 1$ dynamical
ensembles generated by the Hadron Spectrum Collaboration. The use of novel techniques and a 
large basis of interpolating operators have allowed us to extract these spectra to a 
high degree of statistical precision, extract states of high spin and observe candidate
hybrid mesons. We interpret and discuss our results in light of the current experimental 
situation.}

\FullConference{Xth Quark Confinement and the Hadron Spectrum\\
                 8-12 October 2012\\
                 TUM Campus Garching, Munich, Germany}

\begin{document}

\label{sec:intro}
\section{Introduction}
The observation of the enigmatic $D^{*}_{s0}(2317)^{\pm}$ and $D_{s1}(2460)^{\pm}$ states 
at BABAR \cite{BABAR} and CLEO \cite{CLEO} along with the discovery 
of many unexpected states in the hidden-charm sector, has forced charm physics back 
into the theoretical and experimental spotlight. The nature of these states is unknown 
since they do not fit well with current theoretical understanding, i.e. they do not 
match, or in some cases fit into, the pattern expected by quark potential models \cite{Godfrey,Close}. 
Some of the states have been suggested to be molecular mesons, hybrid mesons (in which 
the gluonic field is excited) or tetraquarks but the emergence of a common 
consensus on the nature of any of these states seems unlikely without further work 
from both experimentalists and theorists.

On the experimental side, BESIII along with the planned PANDA experiment at GSI/FAIR will explore the
charm sector in detail, while on the theoretical side, lattice field theory, 
due to its ab initio nature, is perfectly placed to comment on these unexplained 
states. For example, we have recently studied the charmonium sector via the 
extraction of a highly excited spectrum. The results of this study are presented 
in Refs. \cite{Liu,Liua}. In this proceeding we focus on the open-charm sector, and 
present highly excited spectrum in both the charm-light and charm-strange cases.

The rest of this paper is organised as follows: in section~\ref{sec:lattice} we give our 
computational details along with a brief discussion of our operator construction, while also
explaining the details of our analysis and spin identification 
scheme. In section~\ref{sec:results} we present the charm-light and charm-strange spectra,
interpret our results and compare with experiment.

\section{Lattice calculation and analysis}\label{sec:lattice}
The calculations presented here make use of the anisotropic ensembles generated
by the Hadron Spectrum Collaboration~\cite{Edwards,Lin}, in which there are two 
degenerate dynamical light quarks and a dynamical strange quark ($N_{f} = 2 + 1$).
We use a discretisation in which the spatial lattice spacing, $a_{s}$, and 
the temporal lattice spacing, $a_{t}$, are related via 
$\xi = a_{s}/a_{t} \sim 3.5$. This ensures that we simulate with 
$a_{t}m_{c} \ll 1$ and that the standard relativistic formulation of 
fermions can be used.

The gauge sector is described by a Symanzik-improved anisotropic action 
with tree-level tadpole improved coefficients while the fermionic 
fields are described by a tree-level tadpole improved 
Sheikholeslami-Wohlert anisotropic action with stout-smeared spatial
links~\cite{Edwards,Lin}. The same action is used for the charm quark
as for the light and strange quarks except that the parameter 
describing the weighting of spatial and temporal derivatives in the 
lattice action is chosen to give a relativistic dispersion relation
for the $\eta_{c}$ meson as described in Ref.~\cite{Liu}.

\begin{table}[b]
\begin{center}
\begin{tabular}{ccccc}
Volume & $m_\pi/$MeV & $N_{\rm cfgs}$ & $N_{\rm tsrcs}$ & $N_{\rm vecs}$ \\
\hline
$24^3\times 128$ & 391 & 553 &  16 & 162
\end{tabular}
\caption{The gauge-field ensembles used in this work.  $N_{\rm cfgs}$ and 
$N_{\rm tsrcs}$ are respectively the number of gauge-field configurations and 
time-sources per configuration used; $N_{\rm vecs}$ refers to
the number of eigenvectors used in the distillation method~\cite{Peardon}.}
\label{table:perams}
\end{center}
\end{table}

We set the scale by considering the ratio of the $\Omega$-baryon 
mass measured on these ensembles to the experimental mass.
This corresponds to a spatial lattice spacing of 
$a_{s} \sim 0.12$ fm and a temporal lattice spacing
approximately 3.5 times smaller $a^{-1}_{t} \sim 5.7$ GeV.
Table~\ref{table:perams} summarises the ensembles used in our calculation
with full details given in Refs.~\cite{Edwards, Lin}.

\subsection{Operator construction and spectroscopy on the lattice}
In lattice calculations, spectral information is obtained
via Euclidean two-point correlation functions,
\begin{equation}\label{corr}
C_{ij}(t)= \langle 0|{\cal O}_i(t){\cal O}^{\dagger}_j(0)|0\rangle ~, 
\end{equation}
where ${\cal O}^{\dagger}(0)$ and ${\cal O}(t)$ are the source and
sink interpolating fields respectively. When inserting a complete
set of eigenstates of the Hamiltonian, the correlation
function becomes a spectral decomposition,
\begin{equation}
C_{ij}(t) = \sum_\mathfrak{n} \frac{Z_i^{\mathfrak{n}*} Z_j^{\mathfrak{n}}}{2E_\mathfrak{n}} e^{-E_\mathfrak{n}t} ~,
\end{equation}
where the sum is over a discrete set of states due to the finite 
volume of the lattice. The vacuum-state matrix elements
$Z^{\mathfrak{n}}_i \equiv \langle \mathfrak{n} | \mathcal{O}^{\dagger}_i | 0 \rangle$,
are known as \emph{overlaps} and allow for a spin identification
of states as described in section~\ref{sec:analysis}.

In order to maximise the spectral information we can
obtain from correlation functions we use a large
basis of operators. We employ the same derivative based
operator construction scheme as in Ref.~\cite{Dudek:2010}
including operators that contain up to three derivatives. We
apply the \emph{distillation}~\cite{Peardon} smearing
procedure to the quark fields in each operator. This provides
an efficient method to calculate correlation functions with
large bases of operators while also reducing the contamination
of noisy $UV$ modes that do not make a significant contribution
to the low-energy physics we wish to extract.

The lattice breaks three-dimensional rotational symmetry down
to that of the cubic group, $O_{h}$. So instead of an infinite
number of labels on states, $J \geq 0$,
one instead has a finite number of lattice irreps; the five single-cover irreps 
for states at rest are $A_{1}$, $A_{2}$, $E$, $T_{1}$ and $T_{2}$.
States of $J \geq 2$ have their components split across many
lattice irreps. For each lattice irrep, $\Lambda^{P}$ (where $P$ is parity), and 
flavour sector ($D$ and $D_{s}$), we compute an $N \times N$ matrix of 
correlation functions (equation ~\ref{corr}). Here, $N$ is the number of 
operators used within each lattice irrep as shown in Table~\ref{table:ops}.

The extraction of spectroscopic states from these \emph{correlation matrices} will be the subject of 
the next section. 

\begin{table}[b]
\begin{center}
\begin{tabular}{l l|cc}
$\Lambda$ & & $\Lambda^{-}$ & $\Lambda^{+}$ \\
\hline
$A_1$     & & 18   &  18 \\
$A_2$     & & 10   &  10 \\
$T_1$     & & 44   &  44 \\
$T_2$     & & 36   &  36 \\
$E$       & & 26   &  26 \\
\end{tabular}
\caption{The number of operators used in each lattice irrep $\Lambda^{P}$.}
\label{table:ops}
\end{center}
\end{table}

\subsection{Analysis}\label{sec:analysis}

We apply the variational method~\cite{Luscher,Michael} to our correlation matrices 
in order to find the optimal extraction, in the variational sense,
of energies in a given channel. This amounts to solving
the generalised eigenvalue problem,
\begin{equation}
 C_{ij}(t) v^{\mathfrak{n}}_j = \lambda^{\mathfrak{n}}(t,t_0) C_{ij}(t_0) v^{\mathfrak{n}}_{j} ~,
\end{equation}
where an appropriate reference time-slice, $t_{0}$, must be chosen as described in
Ref.~\cite{Dudek:2010}. The energies are determined by fitting
the dependence of the eigenvalues (\emph{principal correlators}), $\lambda^{\mathfrak{n}}$, on
$(t - t_{0})$. The eigenvectors $v^{\mathfrak{n}}$ are related to the
operator-state overlaps, $Z$, and hence play a vital role
in the spin identification of states which is not as straight
forward as one might expect.

In principle, the spin of a single-hadron state can be determined by extracting the spectrum
at various lattices spacings and then extrapolating to the continuum limit. There, where
full rotational symmetry is restored, energy degeneracies between different lattice
irreps should emerge. However, there are difficulties with this procedure. First, this requires high precision
calculations with successively finer lattice spacings and so with increasing computation cost;
this is not practical with currently available resources. Second, the continuum spectrum
can exhibit physical near degeneracies and the question arises as to how to identify, without
infinite statistics, which degeneracies are due to the lattice discretisation and which are
physical degeneracies.

To circumvent these problems we use the spin identification scheme
described in Refs.~\cite{Dudek:2010,Dudek:2009} which can be used at
a single finite lattice spacing. Lattice operators respect the symmetry
of the lattice but they also carry a `memory'
of the continuum spin, $J$, from which they came. If our
lattice is reasonably close to restoring continuum symmetry, at least
on the hadronic scale, then we expect that an operator coming from
a continuum spin $J$ will overlap predominantly onto states of
continuum spin $J$~\cite{Davoudi}. This has been shown to be the case
on these ensembles in many previous studies, for example Ref.~\cite{Liu}.
We hence use the operator-state
overlaps to preform a spin identification of each extracted state
in a given irrep, and use the pattern of the $Z$ values to
recombine components of $J \geq 2$ states since components
of the same $J$ state spread across different irreps have
the same $Z$ values up to discretisation effects. In order to quote
masses for the $J \geq 2$ states we preform joint fits to the
principal correlators in each irrep as described in Ref.~\cite{Dudek:2009}.

\section{Results}\label{sec:results}

\begin{center}
\begin{figure}[t]
\includegraphics[width=1.0\textwidth]{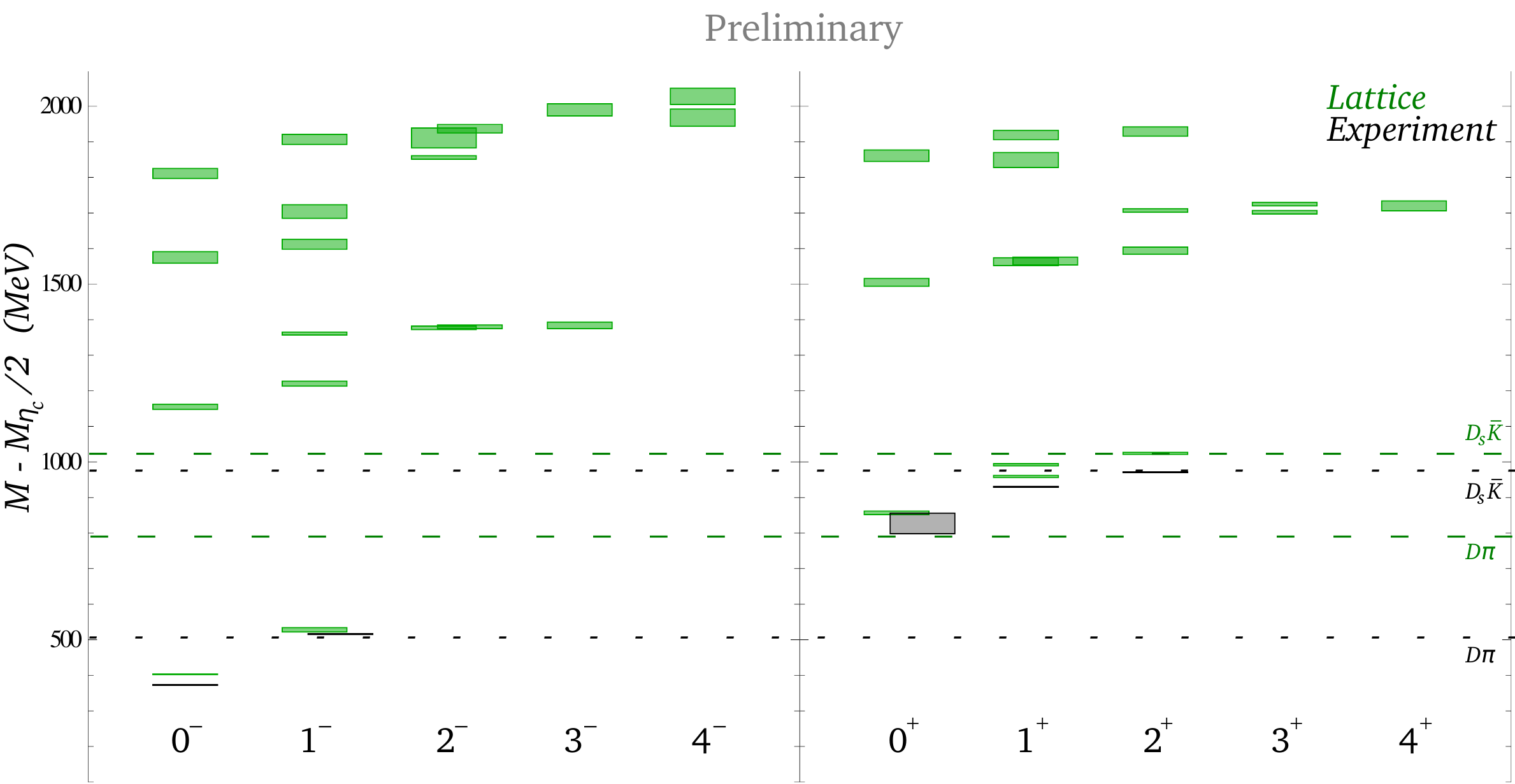}
\caption{The charm-light ($D$) meson spectrum up to around $3.8$ GeV labelled by $J^{P}$. The green boxes are our calculated masses while the black boxes correspond to experimental masses of neutral charm-light mesons from the PDG summary tables~\cite{PDG:2012}. We present the calculated (experimental) masses with half the calculated (experimental) $\eta_{c}$ mass subtracted. The vertical size of each box indicates the one sigma statistical uncertainty. The dashed lines show the lowest non-interacting $D\pi$ and $D_{s}\bar{K}$ thresholds, using our measured masses (coarse green dashing) and experimental masses (fine black dashing).}
\label{fig:D_physical}
\end{figure}
\end{center}

\begin{center}
\begin{figure}[t]
\includegraphics[width=1.0\textwidth]{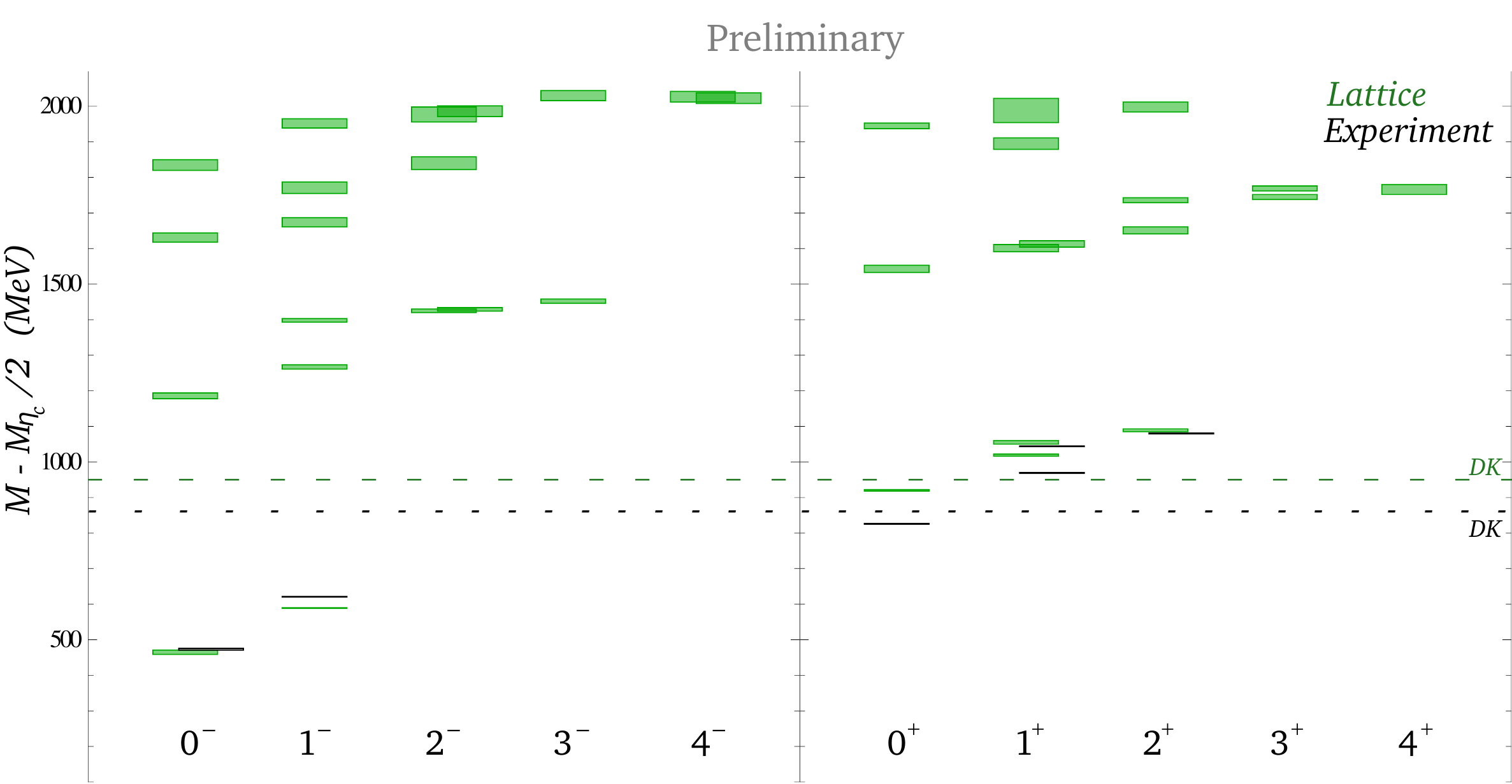}
\caption{As Fig. 1 but for the charm-strange ($D_{s}$) meson spectrum. The dashed lines indicate the lowest non-interacting $DK$ threshold using our measured masses (coarse green dashing) and using experimental masses (fine black dashing).}
\label{fig:Ds_physical}
\end{figure}
\end{center}

The spin-identified charm-light and charm-strange spectra are shown in Figs.~\ref{fig:D_physical} and~\ref{fig:Ds_physical} respectively. The calculated (experimental) masses have had half of the calculated (experimental) $\eta_{c}$ mass subtracted from them in order to reduce the systematic error from the tuning of the bare charm quark mass, which is described in Ref.~\cite{Liu}. In Fig.~\ref{fig:D_physical}, we show the lowest non-interacting $D\pi$ and $D_{s}\bar{K}$ thresholds from both our calculated values (coarse green dashing) and using experimental values (fine black dashing). The lines in Fig.~\ref{fig:Ds_physical} represent the lowest non-interacting $DK$ threshold, again using our calculated values (coarse green dashing) and experimental values (fine black dashing).

\subsection{Interpretation of results and comparison with experiment}

The overlap factors, $Z$, can aid in the interpretation of extracted states as they can be used to identify their structure. As can be seen in Figs.~\ref{fig:D_physical} and~\ref{fig:Ds_physical}, the extracted spectrum of states in the charm-light and charm-strange sectors follow a similar pattern. Most of the states fit into the $n^{2S+1}L_{J}$ pattern expected by quark potential models, where $n$ is the radial quantum number, $S$ is the spin of the quark-antiquark pair, $L$ is the relative orbital angular momentum and $J$ is the total spin of the meson.

In the negative parity sector of both spectra we find an $S$-wave pair $[0^{-}, 1^{-}]$ along with their first and second excitations $\sim 700$ MeV and $\sim 1400$ MeV respectively higher. Also in the negative parity sector, we find a full $D$-wave set $[(1,2,3)^{-}, 2^{-}]$ at $\sim 1400$ MeV, what appears to be parts of an excited $D$-wave set and parts of a $G$-wave set $[(3,4,5)^{-}, 4^{-}]$ at $\sim 2000$ MeV. We do not observe the $5^{-}$ state needed to complete the $G$-wave set but this is to be expected; we only use operators subduced from those that have continuum spins $0 \leq J \leq 4$. Operators that contain four derivatives will posses the required angular structure to access states of spin five.

In the positive parity sector we calculate a full $P$-wave set $[(0, 1, 2)^{+} , 1^{+}]$ around the $D_{s}\bar{K}$ threshold in the charm-light spectrum and around the $DK$ threshold in the charm-strange spectrum. About $600$ MeV higher in both spectra we find the first excitations of the $P$-wave set. We also see a full $F$-wave set $[(2, 3, 4)^{+}, 3^{+}]$ at $\sim 1700$ MeV in both spectra.

\begin{center}
\begin{figure}[t]
\includegraphics[width=1.0\textwidth]{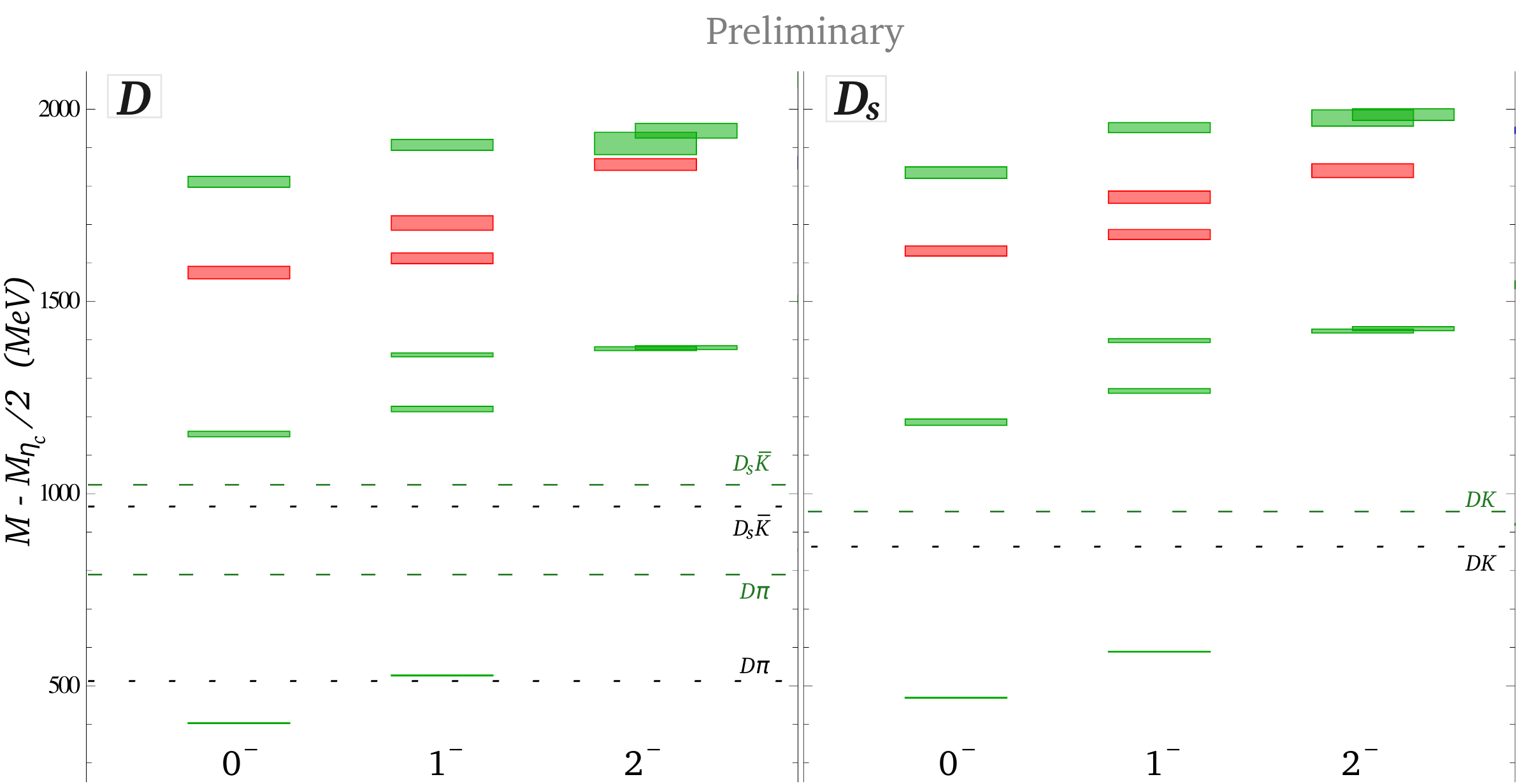}
\caption{The negative-parity charm-light (left panel) and charm-strange (right panel) meson spectra showing only channels where we identify hybrid candidates. The red boxes are identified as states belonging to the lightest hybrid supermultiplet as discussed in the text and other notation is as in Figs. 1 and 2.}
\label{fig:hybrids}
\end{figure}
\end{center}

In the negative parity sector of both spectra we observe four states at $\sim 1600$ MeV that appear to be supernumerary to the pattern of states predicted by quark potential models. Due to their strong overlap with operators proportional to the field strength tensor we interpret these states as the lightest supermultiplet of hybrid mesons and highlight them in red in Fig.~\ref{fig:hybrids}. The pattern of states observed in the supermultiplet is the one expected if a quark-antiquark pair in an $S$-wave configuration is coupled to a $1^{+-}$ quasi-gluon, and its appearance at an energy scale $\sim 1200$ MeV above the lightest quark-antiquark state is in agreement with what was observed in both the light meson sector~\cite{Dudek:2011} and the charmonium sector~\cite{Liu}. 

The four supernumerary states in the positive parity sector $\sim 1500$ MeV above the lightest conventional quark-antiquark state are also interpreted as candidate hybrid mesons, again due to their relatively strong overlap with operators proportional to the field strength tensor. We note, that in the previous studies of Refs.~\cite{Liu,Dudek:2011}, a first excited supermultiplet of hybrid mesons is observed at a similar energy scale, but we do not complete the supermultiplet as other states in this energy region were not robustly determined.     
 
The pattern of our extracted states agree qualitatively with current experimental determinations but some of our states differ quantitatively. The $S$-wave hyperfine splittings we calculate differ by $\sim 20$ MeV from experimental values but this can be explained as an $O(a)$ discretisation effect as in Ref.~\cite{Liu}. In the charm-light sector we find our $P$-wave states heavier than their experimental counterparts which could be due to our unphysically heavy light quarks (our strange quarks are of the correct scale) and/or interaction with the nearby thresholds. In the charm-strange case, two of our $P$-wave states are consistent with experiment but the other two states, which are expected to correspond to the enigmatic $D^{*}_{s0}(2317)^{\pm}$ and $D_{s1}(2460)^{\pm}$, are significantly higher than their experimental counterparts. We note that the $0^{+}$ and $1^{+}$ states are very close to, respectively, the $DK$ and $D^{*}K$ thresholds, and both the experimental and calculated $0^{+}$ states lie the same distance from their appropriate thresholds. This may suggest that the unphysically heavy light quarks are a major contribution to the discrepancy. However, because of the interaction with the threshold, further study is required with multi-hadron operators included in our bases.
  
\section{Summary and outlook}

We have presented preliminary spectra in both the charm-light and charm-strange sectors on $N_{f} = 2 + 1$ dynamical ensembles generated by the Hadron Spectrum Collaboration. The use of a large basis of carefully constructed operators along with a variational analysis has allowed us to extract a large number of states. Our spin identification scheme is crucial to our interpretation as it allows us to identify the $J^{P}$ of not only the low-lying states but also the highly excited states. 

We have identified candidate hybrid mesons due to their relatively strong overlap with operators proportional to the field strength tensor. We observe a pattern of states that is consistent with the lightest hybrid supermultiplet found in previous studies of the light meson~\cite{Dudek:2011} and charmonium sectors~\cite{Liu}.  

In the near future we plan to further study the open-charm sector via calculations along the lines of Refs.~\cite{Dudek:2012,Dudek:2012a}.

\section*{Acknowledgements}

We thank our colleagues within the Hadron Spectrum Collaboration. Chroma~\cite{Scidac} and QUDA\cite{Clark,Babich} were used perform this work on the Lonsdale cluster maintained by the Trinity Centre for High Performance Computing funded through grants from Science Foundation Ireland (SFI), at the SFI/HEA Irish Centre for High-End Computing (ICHEC), and at Jefferson Laboratory under the USQCD Initiative and the LQCD ARRA project. Gauge configurations were generated using resources awarded from the U.S. Department of Energy INCITE program at the Oak Ridge Leadership Computing Facility at Oak Ridge National Laboratory, the NSF Teragrid at the Texas Advanced Computer Center and the Pittsburgh Supercomputer Center, as well as at Jefferson Lab. This research was supported by Science Foundation Ireland under Grant Nos. RFP-PHY-3201 and RFP-PHY-3218. GM acknowledges support from the School of Mathematics at Trinity College Dublin. CET acknowledges support from a Marie Curie International Incoming Fellowship, PIIF-GA 2010-273320, within the 7th European Community Framework Programme.

\end{document}